\documentclass[aps,prx,groupedaddress,12pt,reprint,longbibliography]{revtex4-1}

\usepackage[T1]{fontenc}
\usepackage[ansinew]{inputenc}
\usepackage{amsfonts}
\usepackage{amsmath}
\usepackage{amssymb}
\usepackage[colorlinks,linkcolor=blue,citecolor=blue,urlcolor=blue]{hyperref}
\usepackage{lmodern}
\usepackage{graphicx}
\usepackage{amsmath}
\usepackage{amsthm}
\usepackage{amsfonts}
\usepackage{scalerel}
\def\upar{{\scalerel*{\uparrow}{1}}}
\def\doar{{\scalerel*{\downarrow}{1}}}

\begin{document}

\title{All-Optical Generation and Tuning of  Ultrafast Spin-Hall Current   via  Optical Vortices}

\author{J. W\"{a}tzel}
\email{jonas.waetzel@physik.uni-halle.de}
\author{J. Berakdar}
\email{jamal.berakdar@physik.uni-halle.de}
\affiliation{Institute for Physics,  Martin-Luther-University Halle-Wittenberg, 06099 Halle, Germany}
\date{\today}

%\keywords{Keyword1, Keyword2, Keyword3}

\begin{abstract}
Spin Hall effect, one of the cornerstones in spintronics refers to  the emergence of an imbalance in the spin density transverse to a charge flow in a sample under voltage bias. This study points to a novel way for an ultrafast generation and tuning of a unidirectional nonlinear spin Hall current by means of subpicosecond  laser pulses of optical vortices. When interacting with matter, the optical  orbital angular momentum (OAM) carried by the vortex and quantified by its topological charge is transferred to the charge carriers. The residual spin-orbital coupling in the sample together with confinement effects allow exploiting the absorbed optical OAM for spatio-temporally controlling the spin channels. Both the non-linear spin Hall current and the dynamical spin Hall angle increase for a higher optical topological charge. The reason is the transfer of a higher amount of OAM and the enhancement of the effective spin-orbit interaction strength. No bias voltage is needed. We demonstrate that the spin Hall current can be all-optically generated in an open circuit geometry for ring-structured samples. These results follow from a  full-fledged propagation of the spin-dependent  quantum dynamics on a time-space grid coupled to the phononic environment. The findings point to a versatile  and controllable tool for the ultrafast generation of spin accumulations with a variety of applications such as a source for  ultrafast spin transfer torque and charge and spin current pulse emitter.
\end{abstract}

\flushbottom
\maketitle
\thispagestyle{empty}

\section*{Introduction}

A key issue in spintronics, a field \cite{wolf2001spintronics, vzutic2004spintronics, bader2010spintronics} that utilizes spin dynamics for information processing and storage, has been to tweak the spins by electrical means by applying for instance electric gates or voltage pulses. The promise is a swift and energy-saving operation as compared to using  magnetic fields as well as opening the door for a wider class of materials. A key to electrical manipulation of spin is the spin-orbit interaction (SOI) which can be substantially enhanced for low dimensional structures dictating, for example, new phases at  interfaces~\cite{soumyanarayanan2016emergent, hellman2017interface, schliemann2017colloquium, gambardella2011current}. \\
One of the most intensively researched SOI-induced phenomenona is the spin Hall effect (SHE)~\cite{hirsch1999spin, d1971possibility, dyakonov1971current, kato2004observation, wunderlich2005experimental, sih2005spatial, awschalom2013semiconductor}. SHE has important applications, for instance, it renders possible the generation and detection of spin currents~\cite{ozyilmaz2004current, kiselev2003microwave, jungwirth2012spin, sinova2015spin}, and it contributes with  a spin-torque that can be used to steer magnetization  dynamics~\cite{liu2011spin, miron2011perpendicular, liu2012spin}. The issue of how fast, a possibly non-linear SHE can be triggered raises naturally the question of whether a unidirectional spin Hall current can be generated optically on an ultrafast (THz) timescale. Generically, the electric field component of the optical pulse couples to the carriers orbital motion and the spin is affected via SOI. Photo-induced unidirectional currents, needed for SHE, requires, however, an appropriate break of the  underlying symmetry \cite{awschalom2013semiconductor} which is reflected in a certain topology of the electronic structure posing so constraint on the class of  materials that can be employed. \\
Here we will follow a different route to generate and enhance photocurrent by exploiting the topology of the light fields, meaning the wave front of the optical field (instead of the sample) is assumed to be prepared to have the appropriate symmetry. Below we employ optical vortices that carry and (impart when interacting with matter) a tunable amount of orbital angular momentum (OAM) related to the topological charge of the vortex \cite{allen1992orbital, beijersbergen1993astigmatic, he1995direct, simpson1997mechanical, soskin1997topological, allen2003optical, schmiegelow2016transfer, beijersbergen1994helical}.  {Various methods exist for the generation of optical vortices. To name a few, spatial light modulators (SLM) \cite{ngcobo2013digital, ostrovsky2013generation} and (optimized) spiral phase plates \cite{beijersbergen1994helical, chong2015polarization} were utilized as well as metasurfaces \cite{hakobyan2016tailoring} allowing winding numbers $m_{\rm OAM}>10$ for a wide range of parameters. Various techniques have advantages and limitations: The usage of  SLM allows a dynamical control of the generated OAM light, while the efficiency is relatively low and the beam quality is restricted  by the pixel size of the used nematic liquid crystal cells. In contrast,  spiral phase plates and metasurfaces generate accurate optical vortices but are static approaches. Recent advances in nano-fabrication and engineering of optical materials, nano-scaled integrated OAM laser devices offer versatile tools for shaping the  beams to carry well defined and adjustable OAM \cite{cai2012integrated}. }\\
In  the context of this work, experiments on solid (semi-conducting) samples evidence the light-matter OAM transfer \cite{noyan2015time, shigematsu2016coherent, emile2014electromagnetically}. Below, we demonstrate  how  tuning the optical OAM  enhances the photocurrent and effectively increases he residual SOI (we focus on intrinsic contributions to SHE).
\begin{figure*}[t!]
\centering
\includegraphics[width=12.0cm]{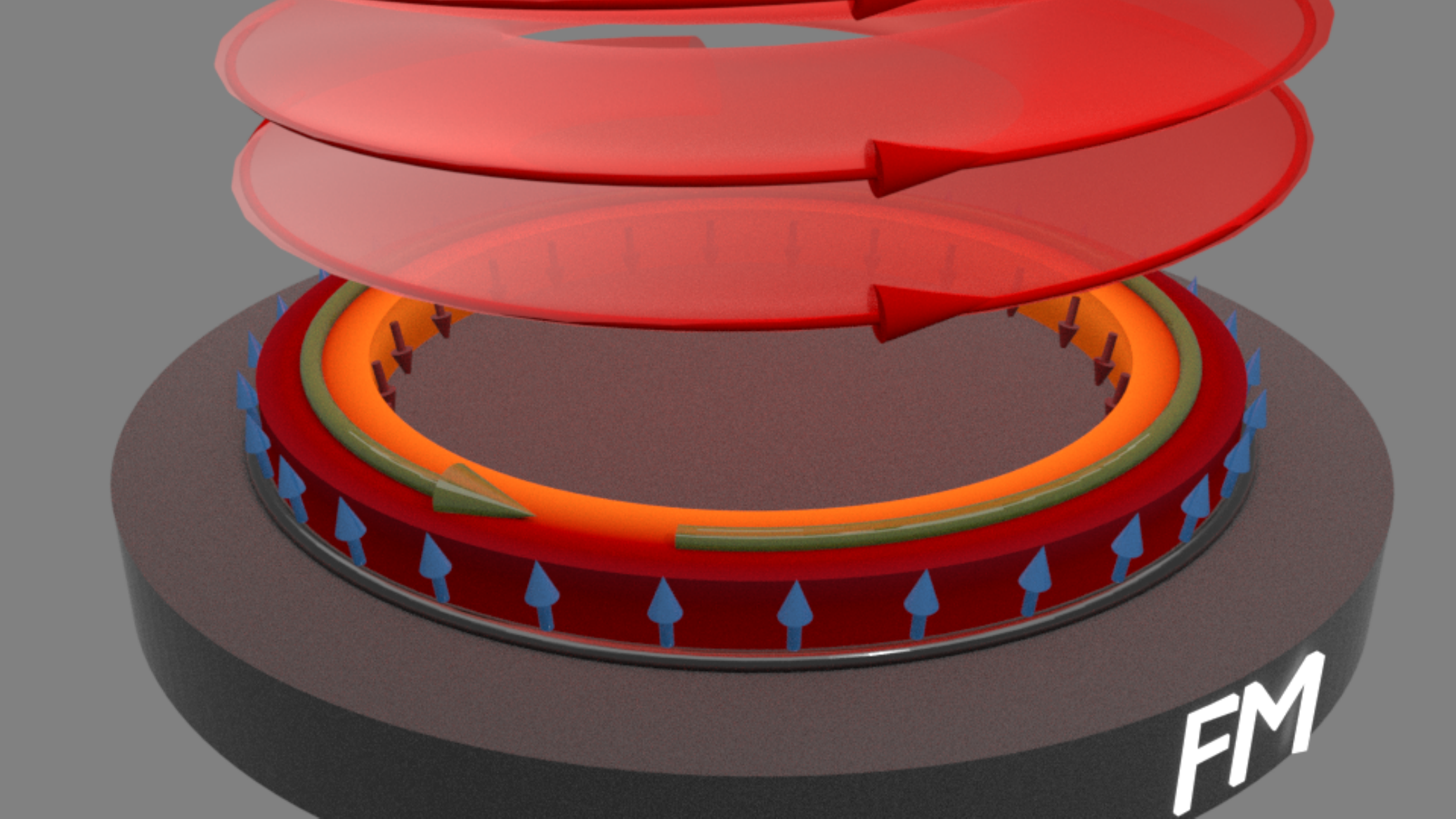}
\caption{Schematics of the optically induced unidirectional spin Hall current. A focused THz optical vortex beam (red circles) initiates a unidirectional circulating current $\pmb{j}$ (green arrow) upon transfer of optical OAM to charge carriers. Due to SOI, the angular orbital motion is accompanied by an drift spin-polarized (Hall) current leading to accumulation of spin-polarized charge density (indicated by arrows) at the ring boundaries which can act with a torque on an interfaced magnet.}
\label{fig1}
\end{figure*}
For all-optical devices, we generate the photo-current in an open circuit geometry by employing a ring structure, as depicted in Fig.\ref{fig1}. The optical vortex whirls the charge density around the ring in a steady-state with a well-defined direction set by the sign of the vortex topological charge. Assisted by SOI, the photocurrent results in a substantial, robust SHE with an associated spin torque that can be, for instance, used to influence the magnetic dynamics in an adjusted magnetic layer, as schematically illustrated in Fig.\ref{fig1}. The effect can be triggered and controlled on a picosecond time scale, and its magnitude and direction are tunable with the OAM of the optical vortex.

\section*{Results}

\subsection*{Setup and theoretical modelling}

For the emergence of photo-induced SHE the SOI is crucial. For considered systems, the Dresselhaus SOI arises from the crystal lattice inversion asymmetry \cite{dresselhaus1955spin} while the Rashba SOI is induced by the structure inversion-asymmetry \cite{rashba1960properties} and depends strongly on the electrical environment, for instance, its effective strength can be tuned by a gate voltage \cite{grundler2000large}. Previous studies considered the time-dependent charge polarization in low-dimensional nano-structures  \cite{vagner1998magnetic, cherkaskiy1999hyperfine, pershin2005laser, sheng2006spin, rasanen2007optimal, kuan2007spectral, zhu2008photoinduced, liu2014spin, joibari2014light, joibari2014light, molnar2005magnetoconductance, foldi2006quantum, souma2005spin,watzel2016centrifugal, watzel2017ultrafast}, to our knowledge, however, spin current generation with optical vortices has not been studied so far.
%Based on this mechanism, we present here a concept for photoinducing the spin Hall effect in a two-dimensional semiconductor quantum ring by optical vortices \cite{allen1992orbital, beijersbergen1993astigmatic, beijersbergen1994helical, he1995direct, simpson1997mechanical, soskin1997topological, allen2003optical} which can be produced in the THz-regime for a wide range of parameters \cite{he2013generation}.
% {Closely related is the photo-generated charge accumulation in quantum rings due to interaction with OAM carrying structured light fields} \cite{watzel2016centrifugal, watzel2017ultrafast}.  {Hence, the optical vortex induced SHE results from a subtle interplay between the transfer of OAM, SOI and finite size effects.}
To be specific, we consider in Fig.\,\ref{fig1} the intra-conduction band carrier dynamics in an appropriately doped InGaAs-InAs-InAlAs-based ring \cite{levy1990magnetization,mailly1993experimental,warburton2000optical} that define the $x-y$ plane. The rings are irradiated by an optical vortex pulse propagating in $z$-direction and tightly focused on the ring. Rashba SOI is substantial \cite{liang2012strong}. The vertical confinement potential $U(z)$ in the growth direction is such that only the lowest subband is involved, meaning  the laser frequency is tuned to the regime $\omega_x<3\hbar^2/(2m^*\Delta h)$ where $\Delta h$ is the ring height in the $z$ direction, and $m^*=0.023m_e$ is the effective mass \cite{averkiev2004energy}. Thus, we focus on the spin-dependent charge dynamics in the $x-y$ plane is the presence of the laser. Adopting polar coordinates $\left\{r,\varphi\right\}$, in the $x-y$ plane the carriers move the azimuthal direction $\hat{\epsilon}_{\varphi}$ while being restricted radially by the  confinement potential $V(r)$. The effective single-particle Hamiltonian reads
\begin{equation}
\hat{H}(t)=\hat{H}_{0} + \hat{H}_{\rm int}(t) + \hat{H}_{\rm ph} + \hat{H}_{\rm B}
\label{eq:Hamiltonian}
\end{equation}
where the unperturbed carriers Hamiltonian $\hat{H}_{0}$, the coupling to the laser fields  $\hat{H}_{\rm int}(t) $, as well as the phononic environment $\hat{H}_{\rm ph} $ and $\hat{H}_{\rm B}$ are given respectively by
\begin{equation}
\begin{split}
\hat{H}_{0} &= \frac{\hat{p}^2}{2m^*} + V(r) + \frac{\alpha}{\hbar}\left[\hat{\sigma}\times\hat{p}\right]_z \\
\hat{H}_{\rm int}(t) &= \frac{e}{2m^*}\left[\hat{\pmb{p}}\cdot\pmb{A}(\pmb{r},t) + \pmb{A}(\pmb{r},t)\cdot\hat{\pmb{p}} + e\pmb{A}(\pmb{r},t)^2\right] \\
&\quad + \frac{e\alpha}{\hbar} \left[\hat{\sigma}\times\pmb{A}(\pmb{r},t)\right]_z + e\Phi(\pmb{r},t) \\
\hat{H}_{\rm bath} &= \sum_{\pmb{q},\lambda} \hbar\omega_{\pmb{q},\lambda}b^{\dagger}_{\pmb{q}\lambda}b_{\pmb{q}\lambda} \\
%\hat{H}_{\rm ph} &= \sum_{\pmb{q},\lambda}M_{\lambda}(\pmb{q})\left(b^{\dagger}_{\pmb{q}\lambda} + %b_{\pmb{q}\lambda}\right)\exp(i\pmb{q}\cdot\pmb{r}) \\
\hat{H}_{\rm ph} &= \sum_{\pmb{q},\lambda}M_{\lambda}(\pmb{q})\left(b^{\dagger}_{\pmb{q}\lambda} \exp(-i\pmb{q}\cdot\pmb{r}) + b_{\pmb{q}\lambda}\exp(i\pmb{q}\cdot\pmb{r})\right) \\
\end{split}
\end{equation}
Here $\hat{\pmb{p}}=-i\hbar\pmb{\nabla}$ is the momentum operator, $\alpha$ is the Rashba SOI strength, $e$ is the elementary charge and $\hat{\sigma}$ is the vector of Pauli matrices. The vortex  vector potential $\pmb{A}(\pmb{r},t)$ fulfills the Lorenz gauge condition \cite{allen1992orbital} $\pmb{\nabla}\cdot\pmb{A}(\pmb{r},t)+(1/c^2)\partial_t\Phi(\pmb{r},t)=0$ where $\Phi(\pmb{r},t)$ is the electric scalar potential that has to be included in the considerations for focused optical vortices. This particular type of spatio-temporally inhomogeneous light-matter interaction is crucial for the emergence of a photo-induced unidirectional spin current. Besides, we will be dealing with short pulses accounting for all multiphoton processes which allow the generation of a substantial spin-current carrying state population. \\
The interaction between the non-equilibrium electrons  and phonons is described by $\hat{H}_{\rm ph}$ \cite{fujisawa2003electrical, ortner2005energy, piacente2007phonon, fujisawa2002allowed} where $M_{\lambda}(\pmb{q})$ is the scattering matrix element, $\pmb{q}$ is the phonon wave vector, $b^{\dagger}_{\pmb{q}\lambda}$ and $b_{\pmb{q}\lambda}$ the phonon annihilation and creation operators, respectively. The parameter $\lambda$ denotes the polarization index. The environmental phonon bath is governed by the Hamiltonian $\hat{H}_{\rm bath}$ in thermal equilibrium with the occupation being characterized by the Bose-Einstein distribution function $n(\pmb{q})=n^0(q)$. Since we are interested in low-lying excitations near the Fermi level we account for interactions with acoustic-phonons only \cite{fujisawa2002allowed, fujisawa2003electrical}, while the excitation energy of optical phonons is above 30\,meV \cite{lockwood2005optical}. In the considered InGaAs-InAs-InAlAs semiconductor quantum ring, an excited electron interacts with the longitudinal acoustic (LA) phonon modes through a deformation potential and with both the longitudinal and the transverse acoustic (TA) phonon modes through piezoelectric fields \cite{piacente2007phonon, price1982electron, mendez1984temperature}. Therefore, the
total scattering matrix element is $\left|M(\pmb{q})\right|^2 = \left|M^{\rm DF}_{\rm LA}(\pmb{q})\right|^2 + \sum_{\lambda={\rm LA,TA}}\left|M_{\lambda}^{\rm PZ}(\pmb{q})\right|^2$.\\
The electron-phonon interaction is spin independent  affecting the spin however when assisted by the SOI. Therefore, the treatment involves both orbital and spin relaxation due to the interaction with phonons. By orbital relaxation, we mean the transition from the excited orbital states (around the Fermi level $E_F$) to all lower states. The spin relaxation is a result of the coupling of the electron spin to the phonons due to SOI and depends on the non-equilibrium level splitting through the action of the optical vortex on the spin states, characterized by the term $\frac{e\alpha}{\hbar}\left[\hat{\sigma}\times\pmb{A}(\pmb{r},t)\right]_z$. Usually, the orbital relaxation is much faster than the spin decay which is in the range of $\mu$s or longer \cite{golovach2004phonon, stano2005spin, stano2006theory, zipper2011spin}.\\
In the case of pure Rashba coupling and confinement potential with circular symmetry, the equilibrium Hamiltonian $\hat{H}_0$ commutes with the total angular momentum operator and its $z$-projection $\hat{J}_z=\hat{L}_z + \hat{S}_z$ where $\hat{S}_z=\frac{\hbar}{2}\sigma_z$ is the operator of the $z$ component of the spin and $\hat{L}=-i\hbar(\pmb{r}\times\pmb{\nabla})$ is the operator of the orbital angular momentum. The form of $\hat{J}_z$ reveals that its eigenvalues $j_z$ are doubly degenerete, because an eigenvalue $\ell_z$ of $\hat{L}_z$  and $1/2$ of $\hat{S}_z$ leads to $j_z = \ell_z + 1/2$ which gives the same result as eigenvalues $\ell_z + 1$ and $-1/2$ of the respective operators.\\
To obtain the ground state spectrum we use the following basis wave functions
\begin{equation}
\phi_{n_{\upar},n_{\doar},\ell}(\pmb{r})=R_{n_{\uparrow},\ell}(r)e^{i\ell\varphi}|\upar\rangle + R_{n_{\downarrow},\ell
+1}(r)e^{i(\ell+1)\varphi}|\doar\rangle
\label{eq:Basis}
\end{equation}
where $\ell$ is the angular momentum quantum number and the radial wave functions satisfy the stationary Schr\"{o}dinger equation $\left[\frac{\hat{p}^2}{2m^*} + V(r)\right]R_{n,\ell}(r)=E^0_{n,\ell}R_{n,\ell}(r)$. The radial quantum number $n$ specifies the radial band. In the case of the used confinement potential $V(r)=a_1/r^2 + a_2r^2 - 2\sqrt{a_1a_2}$ \cite{tan1996electron} the radial states can be found analytically while the energy dispersion is given by $E^0_{n,\ell}=\left(n + \frac{1}{2} + \frac{1}{2}\sqrt{\ell^2 + \frac{2m^*a_1}{\hbar^2}}\right)\hbar\omega_0-\frac{m^*}{4}\omega_0^2r_0^2$ where $\omega_0=\sqrt{8a_2/m^*}$ and $r_0=(a_1/a_2)^{1/4}$. The usage of the basis defined in Eq.\,\eqref{eq:Basis} is advantageous  when calculating the matrix elements $\langle\phi_i|\hat{H}_0|\phi_j\rangle$ (we used the shorthand notation $i=\left\{ n_{\upar},n_{\doar},\ell \right\}$ and separated the radial and angular degrees of freedom) \cite{pietilainen2006energy, szafran2009selective, kushwaha2008magneto}. The resulting matrix is block-diagonal where every individual block corresponds to a specific (orbital) angular quantum number $\ell$ and has the size $2(n_{\rm max}+1)\times2(n_{\rm max}+1)$ where $n_{\rm max}$ is the number of considered radial bands.
\begin{figure*}[t!]
\centering
\includegraphics[width=13.0cm]{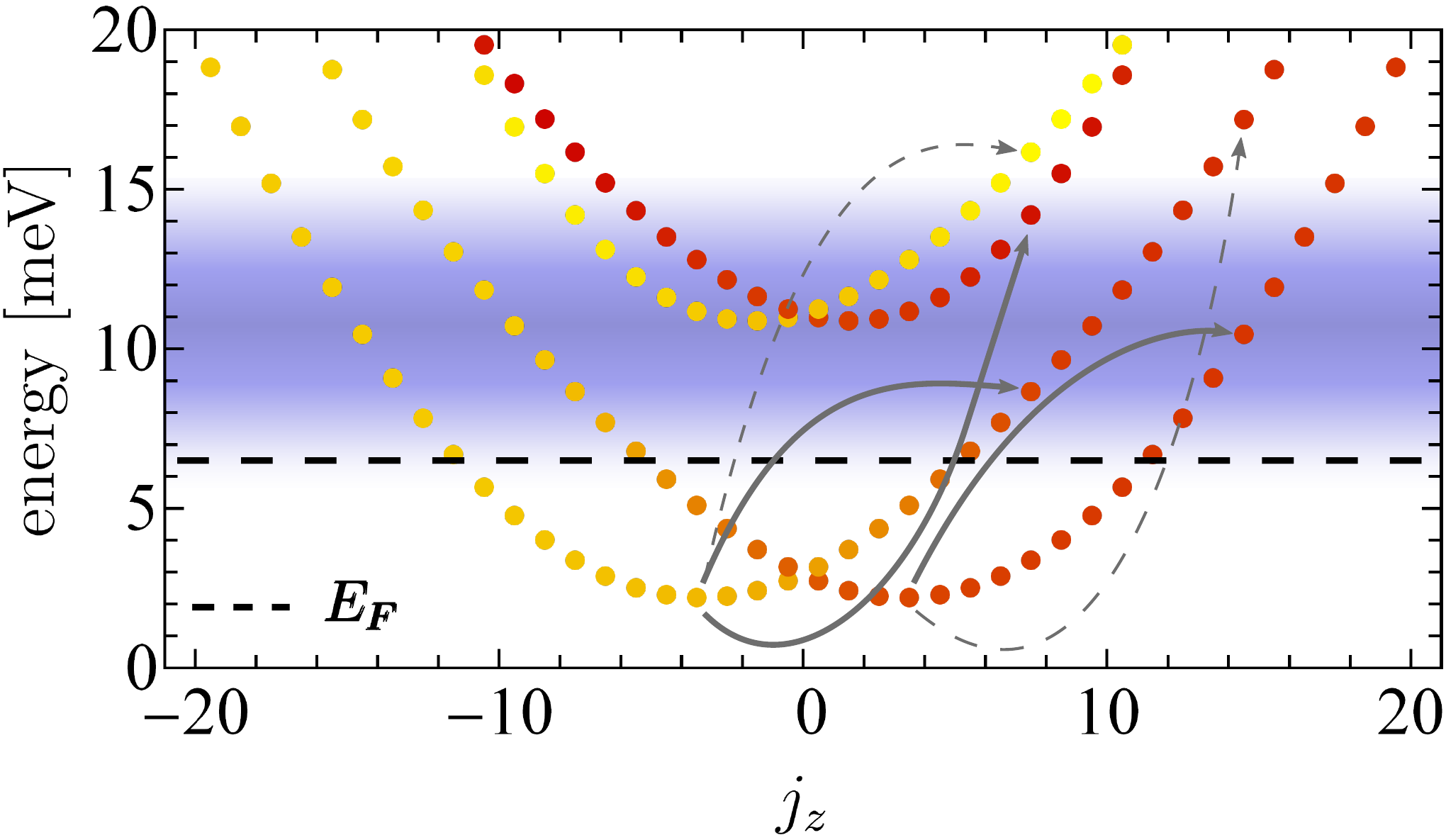}
\caption{Stationary energy  spectrum of the quantum ring with Rashba SOI.  The states are labeled by the eigenvalues of the $z$ component of the total angular momentum $j_z=\ell_z+s_z$ (color gradient  indicates from yellow for $\langle\sigma_z\rangle<0$ through red for $\langle\sigma_z\rangle>0$). $E_F$ is Fermi level. The blue horizontal bar marks the spectral width of the four-cycle optical vortex pulse with photon energy $\hbar\omega_x=8.5$\,meV and topological charge $m_{\rm OAM}=12$, excitations beyond this region are also possible via multiphoton processes.
The thick grey arrows indicate probable transitions pathways for moderately intense laser while the dashed arrows highlight excitations with very low probability. The radial potential of the quantum ring with a radius $r_0=50$\,nm has a confinement strength $\hbar\omega_0=8.5$\,meV while the SOI strength is $\alpha=30$\,meV\,nm corresponding to $\hbar\omega_R=2\alpha/r_0=1.2$\,meV.}
\label{fig2}
\end{figure*}
Therefore, every block can be diagonalized individually. Finally, the resulting equilibrium electron states can be expressed as
\begin{equation}
\begin{split}
\Psi_{n,j}^{(\pm)}(\pmb{r})  =& \sum_{n_{\upar}=0}^{n_{\rm max}}a_{n_{\upar},\ell}^{n,(\pm)}R_{n_{\upar},\ell}(r)e^{i\ell\varphi}|\upar\rangle \\
&+ \sum_{n_{\doar}=0}^{n_{\rm max}}b_{n_{\doar},\ell}^{n,(\pm)}R_{n_{\doar},\ell+1}(r)e^{i(\ell+1)\varphi}|\doar\rangle.
\end{split}
\label{eq:SObasis}
\end{equation}
A typical energy spectrum for a quantum ring with Rashba SOI is shown in Ref.\,\ref{fig2}. It reveals doubled parabolic-like energy curves representing different radial modes $n=0,1,2$ which also characterizes the number of nodes of the corresponding wave functions. The two states corresponding to the same radial mode $n$ and value $j_z=\ell+s_z$ are orthogonal if we take the spin degree of freedom into account. The resulting eigenenergies and states of $\hat{H}_{0}$ can be labeled as $E_{n,j_z}^{(\pm)}$, $|\Psi_{n,j_z}^{(\pm)}\rangle$ respectively, with $j_z$ and $n$ referring to the spatial degrees of freedom (azimuthal and radial coordinates) while the labels $(+)$ and $(-)$ distinguish between positive  ($\langle\Psi_{n,j}^{(+)}|\sigma_z|\Psi_{n,j}^{(+)}\rangle>0$) and negative spin orientation ($\langle\Psi_{n,j}^{(-)}|\sigma_z|\Psi_{n,j}^{(-)}\rangle<0$). \\
The vector potential of a linearly polarized Laguerre-Gaussian vortex beam in the plane $z=0$ is given by $\pmb{A}(\pmb{r},t)=\hat{\epsilon}A_0f_{m_{\rm OAM}}(r)\Omega(t)\exp\left[i(m_{\rm OAM}\varphi-\omega_xt)\right] + {\rm c.c}$ where $A_0$ is the amplitude, $\hat{\epsilon}=\hat{\epsilon}_x$ the polarization vector, $\Omega(t)=\cos^2(\pi t/T_p)$ the pulse envelope for $t\in[-T_p/2,T_p/2]$, $T_p=2\pi n_{\rm cyc}/\omega_x$ the pulse length, $n_{\rm cyc}$ the number of optical cycles.  {This type of optical vortices characterizes a diffractive light wave: indeed, the primary propagation direction (and therefore the direction of the linear momentum) is in the $z$-direction. However, it consists also of plane waves with a distribution of different transverse momenta with the consequence that $\nabla\cdot\pmb{A}(\pmb{r},t)\neq0$. An important quantity is the topological number $m_{\rm OAM}$ which characterizes the longitudinal component of the (carried) orbital part of the angular momentum as an intrinsic property of the optical vortex \cite{Bliokh2015transverse}. Optical vortices reveal a transverse extrinsic orbital part of the angular momentum which is the origin of the spin Hall effect of light (transverse beam shift) in optical reflections \cite{Merano2010orbital}. We note that the separation in the spin and orbital part of the angular momentum is generally not gauge invariant while the \emph{total} angular momentum is invariant \cite{thide2014physics, bialynicki2011canonical}. The spin and orbital parts of the angular momentum correspond to distinct symmetries of the free electromagnetic field and hence are separately conserved quantities \cite{barnett2016natures}. In the Lorenz gauge, the total (longitudinal) orbital and spin angular momenta $\ell\hbar$ and $\sigma_z\hbar$ for the Laguerre-Gaussian modes are obtained.}
The considered radial distribution function of the Laguerre-Gaussian mode \cite{allen1992orbital} has a radial index $p=0$, i.e. $f_{m_{\rm OAM}}(r)=(\sqrt{2}r/w_0)^{m_{\rm OAM}}\exp(-r^2/w_0^2)$. Note, that the usual normalization constant is incorporated into the laser amplitude $A_0$ and we used that the generalized Laguerre polynomials $L_{p=0}^{m_{\rm OAM}}(x)=1$. Using the Lorenz gauge the electric scalar potential is given by $\Phi(\pmb{r},t)=-c^2\int_{-\infty}^t{\rm d}t'\,\nabla\cdot\pmb{A}(\pmb{r},t')$. The intensity profile of the optical vortex beam forms a donut shape focused onto the ring.\\
The interaction of the charge carriers with the optical vortex beam leads to an imbalance in the  internal OAM state with an amount determined by the optical topological charge $m_{\rm OAM}$: $\ell_f=\ell_i + m_{\rm OAM} \pm 1$ \cite{watzel2016centrifugal}.  The initial degeneracy  with respect to clockwise and anti-clockwise angular motion is then broken. The result is the emergence of unidirectional circulating currents (for more details we refer to the supplementary materials).

\subsection*{Numerical results}

We applied our theory to an InGaAs-InAs-InAlAs ring with a radius of $r_0=50$\,nm
with a radial confinement strength  $\hbar\omega_0=8.5$\,meV. The ring width is $\Delta r=25$\,nm. The effective mass of the host material is $m^* = 0.023m_e$ while the effective Rashba SOI strength is tunable between $\alpha = 5$\,meV\,nm and
$\alpha = 40$\,meV\,nm. The central photon energy of the applied  four-cylce optical vortex  pulse is $\hbar\omega_x=8.5$\,meV causing transitions to the second radial band (cf.\,Fig.\,\ref{fig2}) for the appropriate topological charges $m_{\rm OAM}$. For a qualitative comparison between different winding numbers, the amplitude of the vector potential $\pmb{A}(\pmb{r},t)$ has to be normalized to the photon number which will be fixed in the following discussions.
\begin{figure*}[t!]
\centering
\includegraphics[width=13cm]{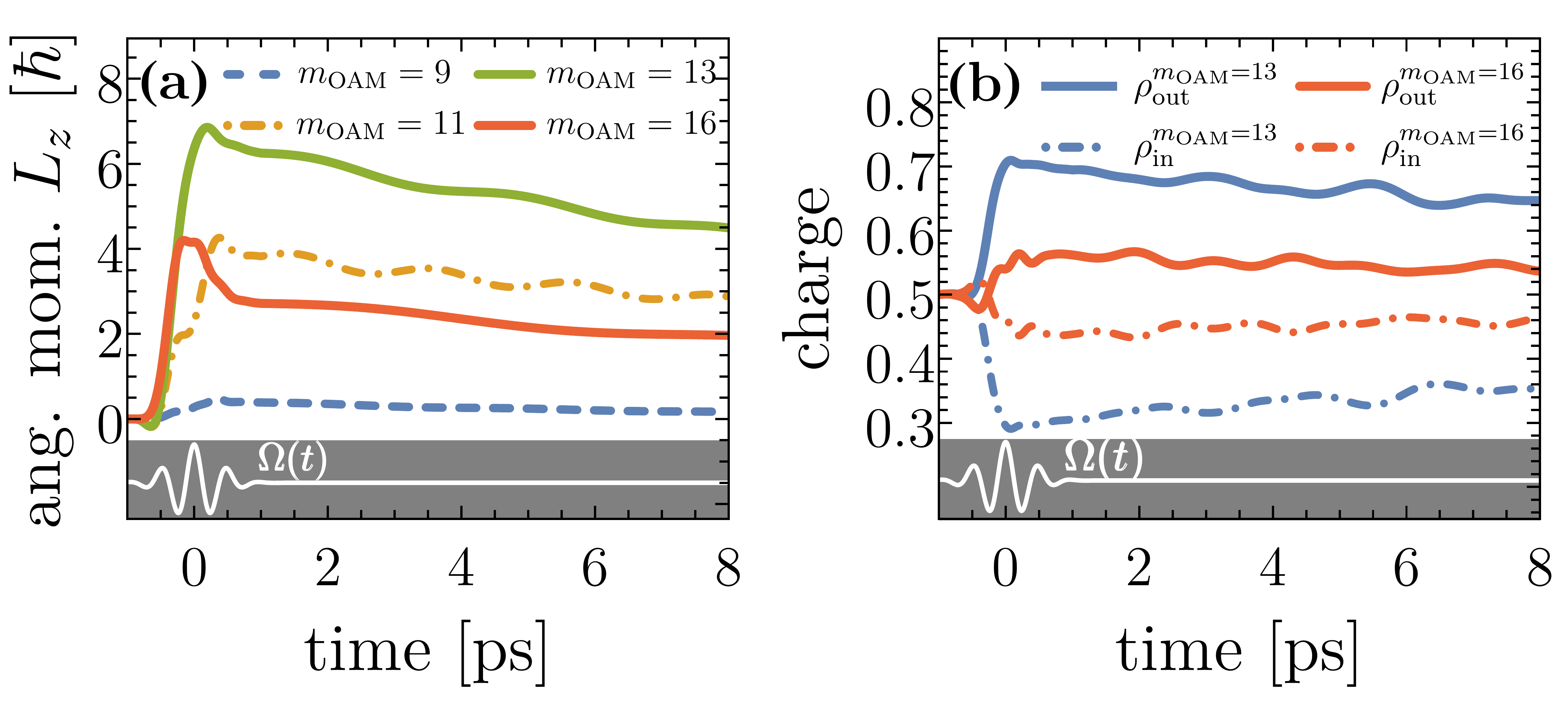}
\caption{Spin-unresolved charge dynamics. (a) The time-dependent expectation value of the orbital angular momentum operator $\hat{L}_z$ (referring to the total quantum ring) for different topological charges $m_{\rm OAM}$ of the irradiated optical vortex beam (the black curve is temporal envelope). The values of $L_z$ reflect the nicely the band width of the electromagnetic pulse. (b) The time-dependent charge accumulation $\rho_{\rm in}(t)$ and $\rho_{\rm out}(t)$ for the inner region $(r<r_0)$ and outer region $(r>0)$ of the quantum ring for two topological charges. Particle conservation is guaranteed since $\rho_{\rm in}(t)+\rho_{\rm out}(t)\equiv1$. The black curve illustrate the temporal behaviour of the four-cycle optical vortex beam. The SOI strength is $\alpha = 30$\,meV\,nm.}
\label{fig3}
\end{figure*}
 { The topological charge of the optical vortex applied here ranges from $m_{\rm OAM}=-20$ to $m_{\rm OAM}=+20$ which is possible with recent experimental techniques \cite{ngcobo2013digital, hakobyan2016tailoring} }. Typically, we consider moderate peak intensities around $I_x=10^{7}$ W/cm$^2$. \\
Fig.\,\ref{fig3}  shows the general influence of the  optical vortex beam imparting  orbital angular momentum to the charge carriers with respect to the $z$ axis. Even for single-photon processes a higher topological charge $m_{\rm OAM}$ leads in general to a larger OAM change $\Delta\ell=m_{\rm OAM}\pm1$ as long as the transition frequency $\omega_{if}=(E_f-E_i)/\hbar$ lies within the band width of the applied pulse which is evidenced by the optical transition amplitudes (see supplementary materials). Therefore, we find a natural limitation of the maximal transferable OAM to the system given by the pulse length $T_p$ which determines the band width. For clarity, in Fig.\,\ref{fig2} the band width of the applied four-cycle optical vortex pulse is shown which also nicely explains which states are involved in the dynamics corresponding to Fig.\,\ref{fig3}(a).  {Here, the dynamical buildup of the $z$-component of the orbital angular momentum $L_z$ of the whole quantum ring is demonstrated which is directly related to the longitudinal component of the light's orbital angular momentum, characterized by $m_{\rm OAM}$. Note, that due to the restriction of the charge dynamics to the $x-y$ plane the in plane components $L_r$ and $L_\varphi$ are zero.}
We find the highest OAM transfer in the case $m_{\rm OAM}=13$ because the associated transition paths belong to the position of the energetic maximum within the band width while, for instance, the case $m_{\rm OAM}=16$ belongs to a drastically reduced photoexcitation probability.
The imparted OAM affects the spin channels via SOI. The excited and non-equilibrium spin-dependent populations interact with the phonon bath leading to irreversible relaxation processes. We find the unidirectional currents are sizable for some picoseconds. A subsequent vortex pulse is needed to build up the current again.\\
In Fig.\,\ref{fig3}(b) the charge accumulation at the ring boundaries is demonstrated which correlates with the gain in OAM in time \cite{watzel2016centrifugal}. Physically, the increase in the angular momentum results in a higher effective centrifugal potential pressing the charge density to the outer ring boundaries, i.e. $\rho_{\rm out}(t)>\rho_{\rm in}(t)$, provided that the quantum confinement is strong enough to avoid a spill out. Here, $\rho_{\rm in/out}(t)=\int_{r\lessgtr r_0}{\rm d}\pmb{r}\langle\pmb{r}|\rho(t)|\pmb{r}\rangle$ where the radial integration is restricted to the domains $r<r_0$ ($\rho_{\rm in}$) and $r>r_0$ ($\rho_{\rm out}$). From Fig.\,\ref{fig3}(b) one can deduce that the strength of this non-equilibrium charge separation is proportional to the amount of the gained OAM during the interaction with the vortex beam. Note that this charge accumulation conserves the particle number since $\rho_{\rm in}(t)+\rho_{\rm out}(t)\equiv1$. The pronounced charge separation decays in time due to relaxation processes as a result of the interaction between the photoexcited electron carriers and the phonon bath. \\
 { Recalling Lorentz reciprocity theorem, one may wonder whether the demonstrated process is reversible, meaning whether a generation of radiation carrying OAM is possible by a spin dependent  loop current.
	At first sight one may anticipate a positive  answer, at least in our case, based on the fact that the charge current in our case is a non-diffusive current (before the phonon path becomes active) and the dynamics should be indeed reversible. Apart from complications related to the spin dynamics reversibility in the presence of spin-orbital coupling, the issue of whether optical vortices (even if generated locally) can then propagate to the far region is not yet settled.  In this connections, we refer to a recent study on the high harmonic emission  of current carrying quantum rings \cite{watzel2017tunable}.  We also note  that an electron in circular motion may emit radiation with helical phase structure hinting  on a carried OAM \cite{katoh2017helical}. }
Along with the photoinduced unidirectional current and charge centrifugal accumulation one may expect some edge spin density accumulations. One note however that the time scale for the charge and spin dynamics are different and it is not clear on which time scale a spin current and spin accumulation will build up and persist.  To gain insight, we consider the time-dependent spin polarization of the inner and outer ring regions as defined by
\begin{equation}
S_{\rm in/out}(t)=\frac{\rho_{\rm in/out}^{\upar}(t)-\rho_{\rm in/out}^{\doar}(t)}{\rho_{\rm in/out}(t)}
\label{eq:spinpolarization}
\end{equation}
where $\rho^{\upar/\doar}_{\rm in/out}(t)=\int_{r\lessgtr r_0}{\rm d}\pmb{r}\langle\pmb{r}|{\rm Tr}\left\{\Lambda_{\upar/\doar}\rho(t)\right\}|\pmb{r}\rangle$ and $\Lambda_{\upar/\doar}$ is the projection operator on the spin-up and spin down states of $\sigma_z$, respectively.  Fig.\,\ref{fig4}  presents the spin polarization of the inner end exterior ring regions for two different optical vortex beam setups. Both regions are initially  spin unpolarized, meaning $S_{\rm in/out}(t\rightarrow-\infty)=0$ in equilibrium. During the laser interaction the 
\begin{figure*}[t!]
\centering
\includegraphics[width=13cm]{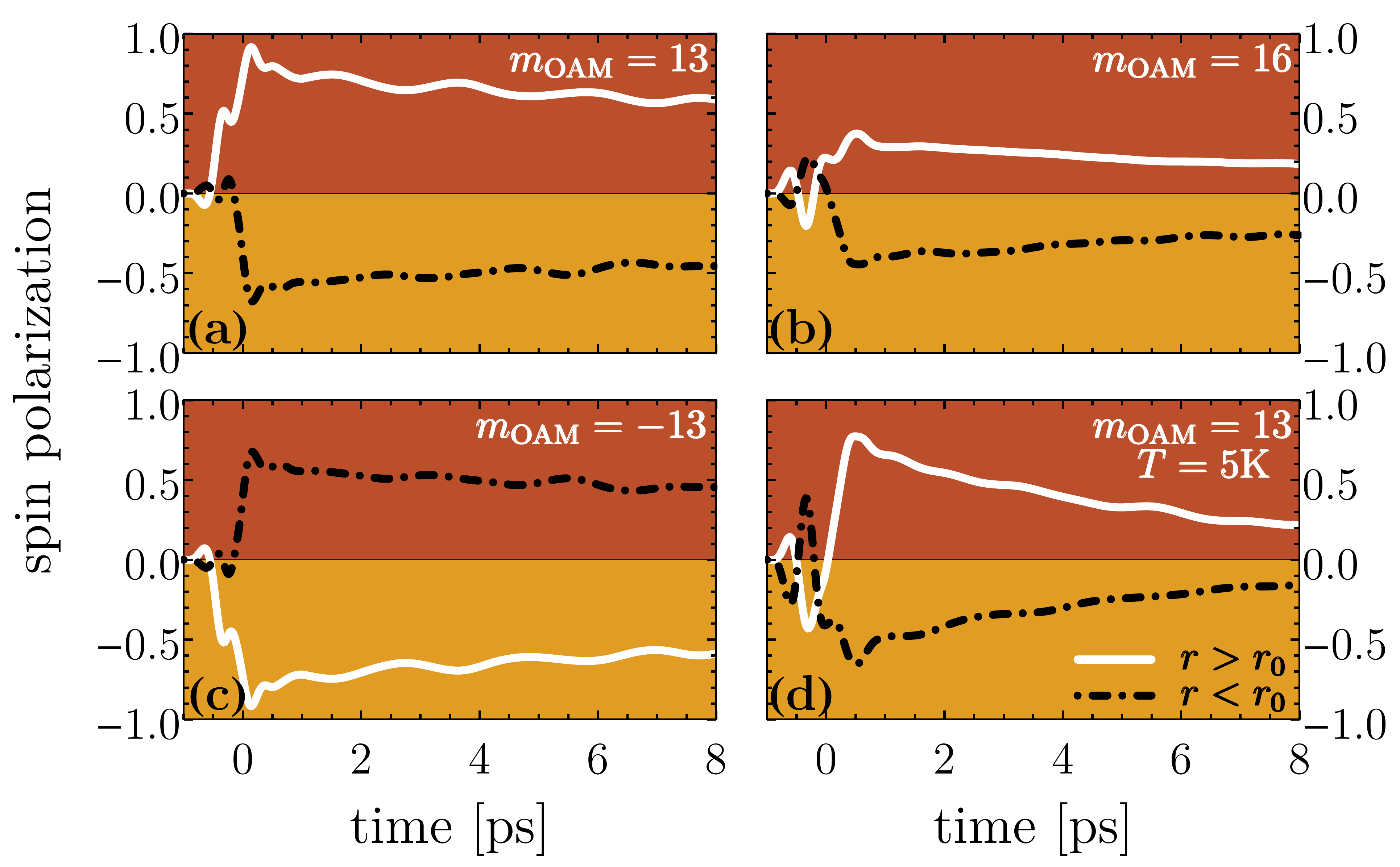}
\caption{The time-dependent spin polarization of the inner $(r<r_0)$ and outer $(r>r_0)$ region of the driven quantum ring for the topological charges (a) $m_{\rm OAM}=13$, (b) $m_{\rm OAM}=16$ (c) $m_{\rm OAM}=-13$ and (d) $m_{\rm OAM}=13$. The pulse and system parameters are the same as in Fig.\,\ref{fig3}. The temperatures for (a)-(c) is 1K and for (d) is 5K.}
\label{fig4}
\end{figure*}
increasing spin separation in the non-equilibrium state builds up. While the charge carriers at the outer ring boundary show more and more a definite positive spin polarization, we find an accumulation of spin-down states at the inner barrier in time implying  a photo-induced spin Hall effect.  Note, $|S_{\rm in}|\neq|S_{\rm out}|$ which reflects the fact that $\left[\hat{H}(t),\sigma_z\right]_-\neq0$ (cf. also Fig.\,\ref{fig2}).
For a positive topological charge an initial state in the spin-up channel $(+)$ ($\langle\sigma_z\rangle>0$, red dots in Fig.\,\ref{fig2}) undergoes a transition (while changing the internal angular momentum  by $m_{\rm OAM}$) to a spin up-channel in the second radial band. In the single photon regime, a spin-flip transition (to a yellow state) is not favored because the corresponding frequency is not within the pulse band width. For a high angular momentum transfer, the charge density is repelled to the outer ring boundary  effectively  spin-polarizing the exterior region. \\
In contrast, for a state in the spin-down channel ($\langle\sigma_z\rangle<0$), yellow dot in Fig.\,\ref{fig2}), a direct spin-flip transition is possible and governed by $\hat{H}_{\rm SOI}(t)=\frac{\alpha}{\hbar}\left[\hat{\sigma}\times\pmb{A}(\pmb{r},t)\right]_z$ as part of $\hat{H}_{\rm int}(t)$ (see supplementary materials) since the corresponding transition frequency is within the band width of the applied beam. {Thus, starting with a spin-down state (yellow dot), the pulse generates a mixture of spin-flip and spin-conserving transitions.}
%{We remark that  the final angular momentum state is smaller in comparison to starting from a electron state with $\langle\sigma_z\rangle<0$ (yellow dot). Thus, due to direct proportionality of the charge drift on the angular momentum state \cite{watzel2016centrifugal} these photoexcited states tend to accumulate at the interior region leading effectively to a negative spin polarization at the inner ring boundary. Second, due to the mix of the spin-flip and spin-orientation conserving transitions the overall spin-down polarization of the inner region is weakened in comparison to the outer boundary.}
Concluding, we predict  the emergence of optical vortex-induced spin Hall drift current due to the transfer of OAM, residual SOI, confinement effects (dictating the initial populations), and temporal properties of the optical vortex beam. \\
The same mechanism applies when changing the sign of the topological charge leading to a switch in the direction of the  spin Hall current, meaning the photo-induced current loop proceeds in the opposite direction leading to negative spin polarization of the exterior domain and an accumulation of spin-up electronic states at the inner boundary, shown in Fig.\,\ref{fig4}(c).
Just like the charge accumulation, depicted in Fig.\,\ref{fig3}(b), also the spin separation depends strongly on the amount of the carried and transferred total OAM to the quantum ring system. In Fig.\,\ref{fig4}(a), the situation for $m_{\rm OAM}=13$ is shown which we identify from Fig.\,\ref{fig3}(a) as the laser beam setup where the nano-structure gains the maximal amount of OAM. Here, we find a strong spin polarization of the inner and outer regions. Around 80 percent spin polarization at the outer ring boundary can be achieved on a ps-time scale just before relaxation becomes sizable. A slight increase of the topological charge to $m_{\rm OAM}=16$ (panel\,\ref{fig4}(b)) results in a distinctly lowered spin separation which comes along with the reduced OAM transfer (cf.\,Fig.\,\ref{fig3}(a)) to the system. The effect of temperatures is illustrated by panel (d) of Fig.\,\ref{fig4} that shows the evolution of $S_{\rm in/out}$ for $T=5$K. Due to the chosen weak confinement and small $E_F$ the initial electronic population is substantially altered and, hence, we find a relatively strong influence of temperatures. However, the general concept does not rely on a specific initial distribution. Rather, it is the residual SOI that in effect converts the optical orbital angular momentum into directed drift spin current.\\
%Very interesting is the result of using a topological charge of $m_{\rm OAM}=11$ which is %shown in panel\,\ref{fig4}(d). Again, the transfer of OAM is reduced because of the %lowered photoexcitation probability but we find that both the inner and the outer regions %of the quantum ring show the same photo-induced spin orientation. This can be explained %by studying once more the electronic structure in Fig.\,\ref{fig2}. By decreasing the %winding number slightly, we find a setup where the spin-flip transition of an electron %from $\langle\sigma_z\rangle<0$ (yellow dot) to the $\langle\sigma_z\rangle>0$ (red dot) %in the second radial band is more probable than the orientation-conserving transition %$\left\{n=0,(-)\right\}\rightarrow\left\{n=1,(-)\right\}$. As a result, practically the %whole quantum ring can be spin-polarized in one direction although the total numbers are %generally smaller than 0.5.\\
Figure\,\ref{fig5}(a) summarizes the dependence on the topological charge of the applied optical vortex beam. For this purpose, let us introduce the time interval $T_{\rm obs}$ straight after the laser pulse which is chosen in a way that the effective relaxation mechanism is not too pronounced meaning $T_{\rm obs}<\tau_{\rm rel}$. Here, $\tau_{\rm rel}$ is the effective relaxation time which is around 25\,ps as revealed by the Redfield tensor in Eq.\,\eqref{eq:RedfieldTensor}. A reasonable choice is $T_{\rm obs}=2\,$ps. As a consequence, the average spin polarizations
\begin{figure*}[t!]
\centering
\includegraphics[width=13cm]{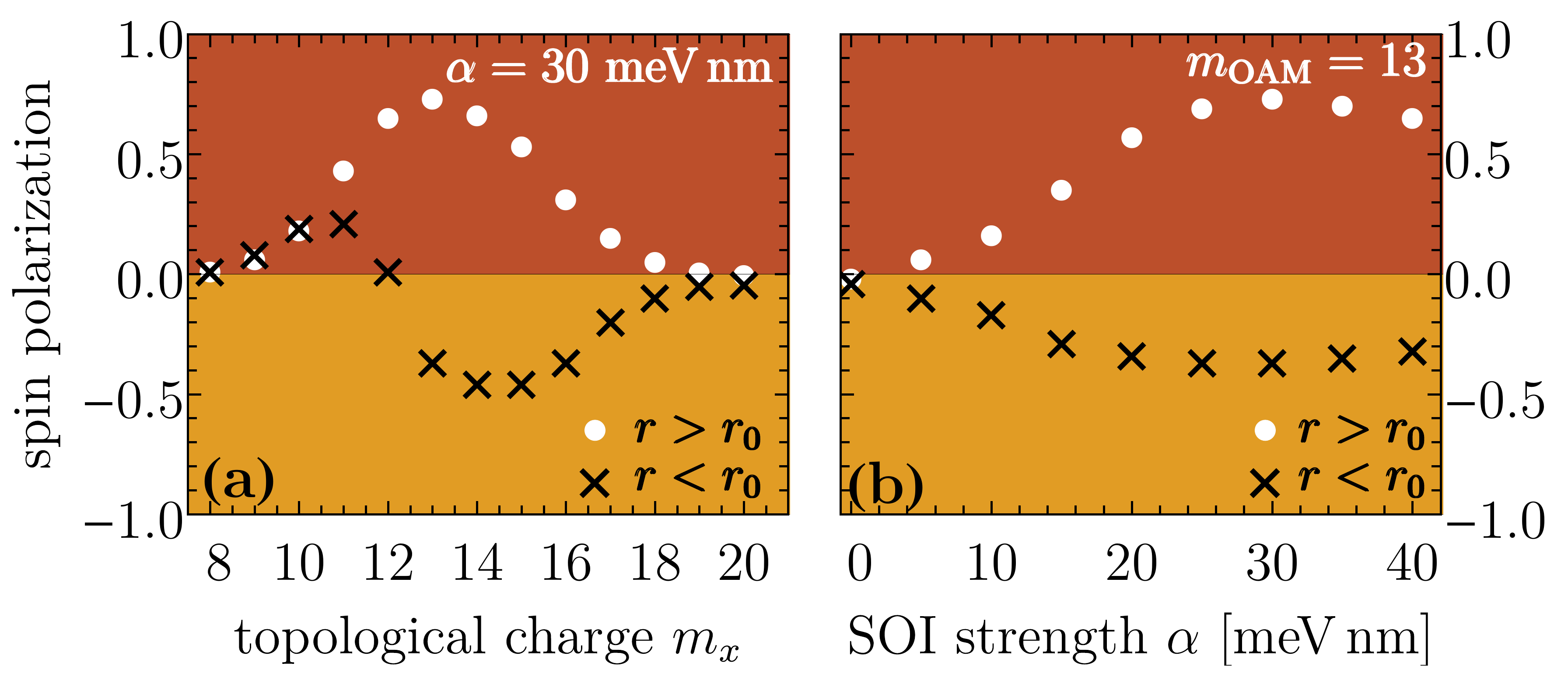}
\caption{(a) The dependence of the spin polarization of the interior and exterior ring regions on the topological charge $m_{\rm OAM}$ of the applied optical vortex pulse. (b) Spin polarizations for different Rashba SOI strengths $\alpha$. The other laser and system parameters are the same as those in Fig.\,\ref{fig2}.}
\label{fig5}
\end{figure*}
\begin{equation}
\overline{S}_{\rm In/Out}=\frac{1}{T_{\rm obs}}\int_{T_p/2}^{T_p/2+T_{\rm obs}} S_{\rm In/Out}(t).
\label{eq:avr}
\end{equation}
are characteristic quantities for the non-equilibrium time regime. As expected we find a maximal spin separation and most pronounced spin polarization in the case $m_{\rm OAM}=13$.  For topological charges below $m_{\rm OAM}=8$ or higher than $m_{\rm OAM}=19$, we find no spin separation due to a vanishing transfer of OAM because the corresponding transition frequencies are beyond the pulse band width, i.e., higher non-linear processes turned out to be very small. An interesting behaviour shows the spin polarization of the inner region which undertakes a transition from a positive spin orientation for smaller $m_{\rm OAM}$ to a spin-down polarization for larger $m_{\rm OAM}$. Thus, the pulse band width together with the amount of transferred OAM (determined by $m_{\rm OAM}$) determines whether spin-conserving or spin-flip transitions occur.\\
Fig.\,\ref{fig5}(b) demonstrate the role of the Rashba SOI strength $\alpha$ for a topological charge $m_{\rm OAM}=13$. Obviously, no spin separation occurs for $\alpha\rightarrow0$. In the domain $\alpha\in[0,30\,{\rm meV\,nm}]$, the spin separation (polarization) enhances with increasing the  Rashba SOI strength. Surprisingly, for coupling strengths above $\alpha=30$\,meV\,nm the spin separation starts decreasing. One can explain this phenomenon with spin crossing effects. Our equilibrium calculations of the electronic structure, depicted in Fig.\,\ref{fig2}, reveal that for $\alpha>20$\,meV\,nm  the spin channels are strongly mixed. Thus, optically induced transitions are accompanied with spin-flips leading so to damping of the overall spin polarization.\\
A key quantity  describing  SHE is the spin Hall angle (SHA) given as the ratio between the transversal spin current and the  longitudinal charge current for moderate bias.
For a dc electric field in the steady state, the SHA is defined in the linear response regime as $\alpha_{\rm SHA}=\sigma^s_{xy}/\sigma_{xx}$, meaning the ratio of the spin Hall conductivity $\sigma^s_{xy}$ and the diagonal term $\sigma_{xx}$ of the conductivity tensor  \cite{sinova2015spin}.
In our case, we deal with time-dependent non-linear processes which are  captured through   the numerical propagation of the density matrix. For a meaningful defintion of SHA in this regime, we use the spin-resolved current density $\pmb{j}^s(\pmb{r},t)={\rm Tr}\left\{\Lambda_s\hat{\rho}(t)\hat{\pmb{j}}\right\}$.  The operator $\hat{\pmb{j}}(\pmb{r},t)=\frac{e}{2m^*} \left[|\pmb{r}\rangle\langle\pmb{r}|\hat{\pmb{\pi}}+\hat{\pmb{\pi}}^\dagger|\pmb{r}\rangle\langle\pmb{r}|\right]$ with $\hat{\pmb{\pi}}=\hat{\pmb{p}}-e\pmb{A}(\pmb{r},t)$ and $\Lambda_s$ projects on the eigenfunctions $|s=\upar,\doar\rangle$ of $\hat{\sigma}_z$. The dynamical, non-linear SHA we quantify by
\begin{equation}
\alpha_{\rm SHA}(t)=\frac{I_r^\upar(t) - I_r^\doar(t)}{I_\varphi^\upar(t) + I_\varphi^\doar(t)}
\label{eq:SHA}
\end{equation}
where $\pmb{I}^s(t)=\int{\rm d}\pmb{r}\,\pmb{j}^s(\pmb{r},t)$. Thus, $\alpha_{\rm SHA}(t)$ describes the ratio between the transverse  \emph{spin-polarized} current $I_r^\upar(t) - I_r^\doar(t)$  (cf.\,Fig.\,\ref{fig4}) and the angular charge current in $\hat{\epsilon}_\varphi$.  \\
In Fig.\,\ref{fig6}(a) the general temperoral characteristics and dependence on the topological charge of the applied optical vortex are shown. The pulse parameters are the same as those used for the results depicted in Figs.\,\ref{fig3}-\ref{fig5}. An increase of the winding number leads to a higher current effectively enlarging the SHA as long as the transition frequency $\omega_{if}$ is in the band width of the four-cycle pulse. In the case of $m_{\rm OAM}=13$ we find the largest SHA which corresponds to the largest acquired angular momentum by the quantum ring structure (cf.\,Fig.\,\ref{fig3}(a)). Panel (b) depicts  the influence of the laser peak intensity  when averaging over the observation interval $T_{\rm obs}$ straight after the laser pulse (cf. Eq.\,\ref{eq:avr}). The results indicate  three regimes: below $5\times10^6$\,W/cm$^2$, the driven quantum ring exhibits a nearly linear-response to laser field strength. For intensities larger than $5\times10^6$\,W/cm$^2$, the SHA increases rapidly signifying non-linearity.  Thus, our results  for $I_x=10^7\,$W/cm$^2$ (Figs.\,\ref{fig3}-\ref{fig5}) are in the non-linear regime. Increasing the intensity further, we observe a saturation effect. Current technical limitations prevent  using higher peak intensities in our numerics as the needed basis set of eigenfunctions increases drastically because multi-photon process become dominant.

\begin{figure*}[t!]
\centering
\includegraphics[width=12cm]{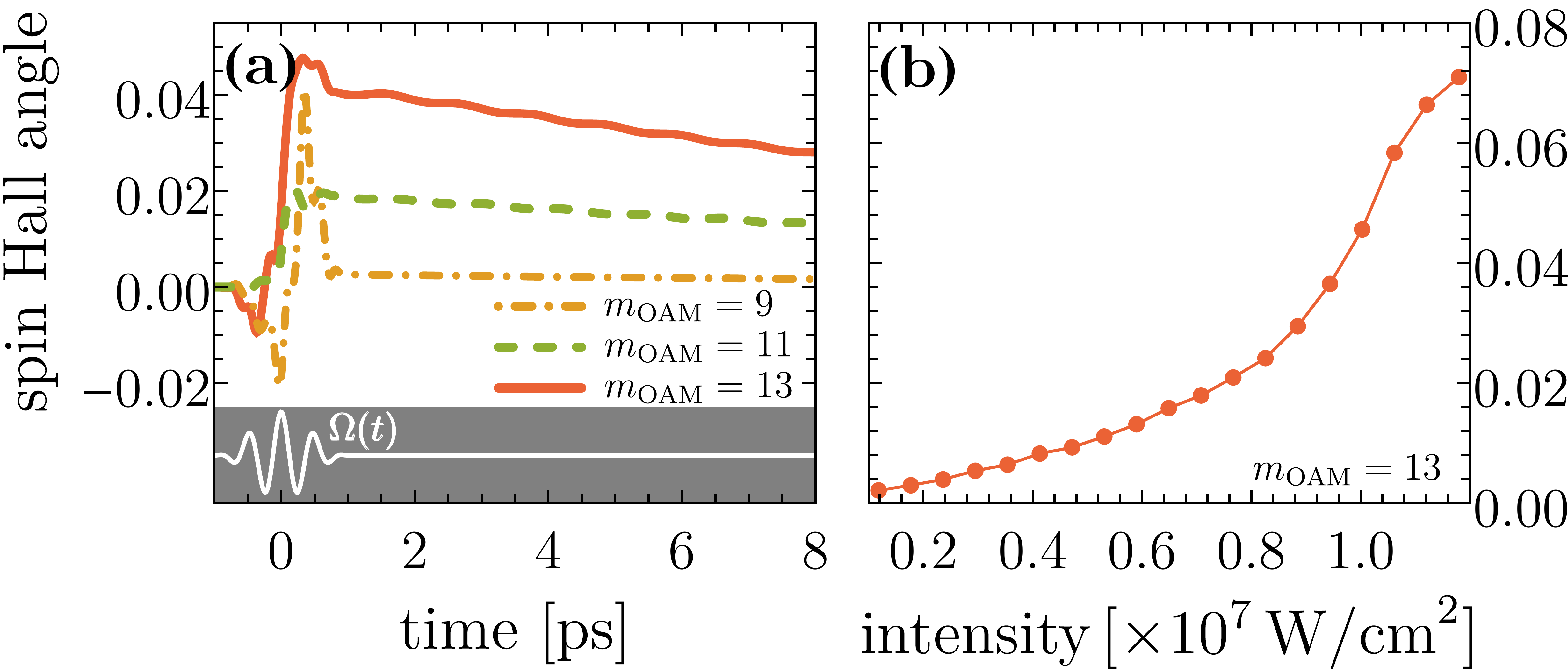}
\caption{ (a) Buildup of  dynamical spin Hall angle  (SHA) for different topological charges $m_{\rm OAM}$ at fixed intensity. An optically-induced orbital angular momentum leads to a higher (photo-induced) charge current and stronger effective spin-orbital effects (cf.\,Fig.\ref{fig4}). (b) Dynamical SHA vs. laser peak intensity.  Linear response is valid for intensities below $5\times10^6$\,W/cm$^2$.}
\label{fig6}
\end{figure*}

\section*{Discussion}

We demonstrated an efficient way for ultrafast photo-induced unidirectional spin Hall current pulses in a two-dimensional semiconductor quantum ring using a few-cycle optical vortex beam. The imparted optical orbital angular momentum ignites a unidirectional looping charge current which, due to the residual spin-orbital coupling induces a drift spin current leading to edge  spin accumulation.
These phenomena can be controlled on the picosecond time scale by the parameters of the pulse, for instance. the magnitude of the spin current can be enhanced by increasing the optical topological charge while keeping the laser intensity unchanged making the method noninvasive.
The mechanism relies on the topology of the light field OAM rather than on the system crystal symmetry and should thus be present for a wide range of structured materials with spin-orbital coupling.
A large number of important applications based on the revealed phenomena can be envisaged.  The triggered boundary spin accumulation is large and can be exploited as a source for an ultrafast spin transfer torque when interfaced  with a ferromagnetic material. Further, when  appropriately contacted with a conductor with a strong spin-orbit scattering  (such as a Pt stripe) the spin density buildup generates a charge current pulse via the inverse spin Hall effect.

\section*{Methods}

In its equilibrium state, the initial density matrix of the quantum ring reads $\rho_{i,j}^0(t\rightarrow-\infty) = f_i^0(E_F,T)\delta_{i,j}$ where $f_i^0(E_F,T)$ is the Fermi-Dirac distribution \cite{moskalenko2006revivals}. The  excitation dynamics of the system is governed by Heisenberg's equation of motion
\begin{equation}
\begin{split}
i\hbar\partial_t\rho_{m,m'}(t)  =& \left[\hat{H}_0 + \hat{H}_{\rm int}(t),\rho_{m,m'}(t)\right]_- \\ &-\sum_{kl}R_{mm'kl}\rho_{kl}(t)
\end{split}
\label{eq:RDM}
\end{equation}
which is space-time-grid propagated in the presence of the vortex  vector potential and includes thus non-linear effects in the external field.
%Here, the first line represents the reversible dynamics which one can interprets as coherent effects in terms of resonant transitions from quantum state $m$ to $m'$ (governed by the transition frequency $\omega_{m,m'}=(E_{m'}-E_m)/\hbar$).
Irreversible open system dynamics is also captured (on the level outlined above). Initially and in equilibrium, the system described by $\hat{H}_0$ and the phonon bath $\hat{H}_B$ are uncorrelated, meaning that the total density matrix is cast as  $\rho_{\rm tot}(t\rightarrow-\infty)= \rho(t\rightarrow-\infty)\rho_B(t\rightarrow-\infty)$ where $\rho_B=e^{-\hat{H}_B/k_BT}/Z_B$ with $Z_B={\rm Tr}\{e^{-\hat{H}_B/k_BT}\}$. Upon excitations the system-bath coupling sets in. We employ the following approximations: the coupling to the phonon bath is weak allowing the use of the Born approximation $\rho_{\rm tot}(t)\simeq\rho(t)\rho_B(t\rightarrow-\infty)$. We assume that the time evolution of $\rho(t)$ depends only on its present value and not on its past state, meaning we use the Markov approximation. Then, the  system (reduced) density matrix $\rho(t)={\rm Tr}_B\left\{\rho_{\rm tot}(t)\right\}$ in Eq.\,\eqref{eq:RDM}  traced over all degrees of freedom of the phonon reservoir $B$  can be obtained \cite{blum1981density, chirolli2008decoherence, mahler2013quantum}. Using standard  density matrix algebra, the Redfield tensor is  expressed as
\begin{equation}
\begin{split}
R_{nmkl} =\delta_{nm}\sum_r\Gamma_{nrrk}^+ + \delta_{nk}\sum_r\Gamma_{lrrk}^- -\Gamma_{lmnk}^+ - \Gamma_{lmnk}^-
\end{split}
\label{eq:RedfieldTensor}
\end{equation}
in terms of transition rates governed by Fermi golden rules expressions
\begin{equation}
\Gamma_{lmnk}^+=\int_0^{\infty}{\rm d}t\,e^{-i\omega_{nk}t}\overline{\langle l|\hat{H}_{\rm ph}|m\rangle\langle n|\hat{H}_{\rm ph}(t)|k\rangle}.
\end{equation}
Here we use that $\Gamma_{lmnk}^+=(\Gamma_{knml}^-)^*$, while $\hat{H}_{\rm ph}(t)=\exp(i\hat{H}_Bt/\hbar) \hat H_{\rm ph}\exp(-i\hat{H}_Bt/\hbar)$. The overbar donates the average over the phonon bath in thermal equilibrium at a temperature $T$ meaning $\overline{\langle l|\hat{H}_{\rm ph}|m\rangle\langle n|\hat{H}_{\rm ph}(t)|k\rangle} = {\rm Tr}_B\left\{\langle l|\hat{H}_{\rm ph}|m\rangle\langle n|\hat{H}_{\rm ph}(t)|k\rangle\rho_B\right\}$. A detailed description of the transition rates is given in the supplementary materials.\\
Crucial for obtaining the transition rates $\Gamma_{lmnk}^+$ is the calculation of all phonon matrix elements $\langle n|\hat{H}_{\rm ph}|m\rangle$ between all states shown in the spectrum in Fig.\,\ref{fig2}. Since we are using spin-resolved eigenstates (Eq.\,\ref{eq:SObasis}) of $\hat{H}_0$, charge and spin relaxation processes due to the spin admixture as a spin-orbit coupling phenomenon are included. For instance, the diagonal elements of $R_{mm'kl}$ reveal the spin relaxation rate which is explicitly given by $T_{1,mm'}^{-1}=2{\rm Re}\left(\Gamma^+_{m,m',m',m} + \Gamma^+_{m',m,m,m'}\right)$ \cite{chirolli2008decoherence}. Note, that the off-diagonal elements of the Redfield tensor can be identified as decoherence rates due to the interaction with the reservoir and are responsible for the time decay of the optical excitations initiated by $\hat{H}_{\rm int}(t)$. Heisenberg's equation of motion in Eq.\,\ref{eq:RDM} is solved using a numerically stable Leapfrog algorithm.
\\
By solving Eq.\,\eqref{eq:RDM} we neglect Coulomb-interactions between the charge carriers.
There are several reasons for this doing: in the stationary case, i.e. $t\rightarrow-\infty$, it can be demonstrated that for a small number of electronic states in a quantum ring with no impurity the inclusion of correlation via Coulomb matrix elements simply shifts the non-interacting spectrum to higher energies \cite{pietilainen1995electron}. Therefore, we calculated numerically various Coulomb matrix elements $V_{abcd}=\langle\Psi_a\Psi_b|V(\pmb{r}-\pmb{r}')|\Psi_c\Psi_d\rangle$ where $V(\pmb{r}-\pmb{r}')=e^2/(4\pi\epsilon_0\epsilon_r|\pmb{r}-\pmb{r}'|)$ and we can confirm that they are actually smaller than the kinetic energy matrix elements.  Thus,  correlation effects lead only to a small perturbation of the underlying electronic spectrum. Importantly, the Coulomb matrix elements conserve the angular momentum, i.e. $m_a=m_c$ and $m_b=m_d$ \cite{zurita2002multipolar}. Since we consider optically induced intra (conduction)band transitions which change the internal angular momentum state, the Coulomb matrix elements between the states which are involved in the electric transition disappear. Consequently, correlation effects play a minor role for the qualitative description of the considered  transitions.
We made sure that the parameters of the pulses are chosen to trigger mainly excitations near the Fermi-level in which case the approximation discussed above remain credible.

\section*{Acknowledgements (not compulsory)}

This work was partially supported by the  DFG through SPP1840 and SFB TRR 227.

\section*{Author contributions statement}

JW and JB contributed equally to the conceptual development, interpretation and data analysis. JW performed the numerical simulations.

\section*{Additional information}

\textbf{Supplementary Information} accompanies this work.\\
\textbf{Competing interests: } The authors declare no competing financial and non-financial  interests.

\end{document}

\title{Supplemental material: All-optical generation and ultrafast tuning of  non-linear spin Hall current}

\author{J. W\"{a}tzel}
\email{jonas.waetzel@physik.uni-halle.de}
\author{J. Berakdar}
\email{jamal.berakdar@physik.uni-halle.de}
\affiliation{Institute for Physics, Martin-Luther-University Halle-Wittenberg, 06099 Halle, Germany}
\maketitle

\section{Dissipative Dynamics}
\label{AppA}
In the low-temperature limit, the interaction between the phonon reservoir $B$ and the quantum ring is written as the product
\begin{equation}
\hat{H}_{\rm ph}=\sum_{\pmb{q}} Q_{\pmb{q}}F_{\pmb{q}}
\end{equation}
where $Q_{\pmb{q}}=M_{\lambda}(\pmb{q})\exp(-i\pmb{q}\cdot\pmb{r})$ is an operator acting on the ring states and
%$F_{\pmb{q}}=\left(b^{\dagger}_{\pmb{q}\lambda} + b_{\pmb{q}\lambda}\right)$
$F_{\pmb{q}}=b^{\dagger}_{\pmb{q}\lambda}$ is a bath operator (we neglect phonon absorption in the low temperature limit). In this case the Redfield-tenstor components are given by \cite{weiss1999quantum}
\begin{equation}
\begin{split}
\Gamma_{lmnk}^+=&\sum_{\nu,\nu'}\langle l|Q_{\pmb{\nu}}|m\rangle\langle n|Q_{\pmb{\nu'}}|k\rangle \\
&\times\int_0^\infty {\rm d}t \langle F_{\pmb{\nu}}F_{\pmb{\nu'}}(t)\rangle_B e^{-i\omega_{nk}t} \\
\Gamma_{lmnk}^-=&\sum_{\nu,\nu'}\langle l|Q_{\pmb{\nu}}|m\rangle\langle n|Q_{\pmb{\nu'}}|k\rangle \\
&\times\int_0^\infty {\rm d}t \langle F_{\pmb{\nu}}(t)F_{\pmb{\nu'} }(0)\rangle_B e^{-i\omega_{lm}t}.
\end{split}
\label{eq:RedfieldComponents}
\end{equation}
The  bath-dependent factors can be expressed by the bath correlation function $\langle F_{\pmb{\nu}}F_{\pmb{\nu'}}(t)\rangle_B={\rm Tr}\left\{F_{\pmb{\nu}}F_{\pmb{\nu'}}(t)\rho_B\right\}$ and $\langle F_{\pmb{\nu}}(t)F_{\pmb{\nu'}}\rangle_B={\rm Tr}\left\{F_{\pmb{\nu}}(t)F_{\pmb{\nu'}}\rho_B\right\}$. A striking property is the decay to zero within a certain time scale, which is set by the bath parameter dependent correlation time $t_c$. Here, the bath time-correlation function is given by \cite{chirolli2008decoherence}
\begin{equation}
\langle F(0)F(t)\rangle_B = \frac{1}{\pi}\int_0^\infty{\rm d}\omega J(\omega)\left[e^{-i\omega t} + 2n^0(\omega)\cos(\omega t)\right].
\label{eq:correlation}
\end{equation}
Here, $n^0(\omega)$ is the Bose-Einstein distribution and
\begin{equation}
J(\omega)=\frac{\pi}{2}\sum_{\pmb{q}} \left|a_{\pmb{q}}\right|^2\delta(\omega-\omega_{\pmb{q}})
\label{eq:Jw}
\end{equation}
is the bath spectral density of the Boson modes $\pmb{q}$. The connection to the system is characterized by the weights $\left|a_{\pmb{q}}\right|^2$ (coupling strength to the electron charges) which are related to the operator $Q_{\pmb{q}}$ by \cite{brandes2002adiabatic,thorwart2001decoherence}:
\begin{equation}
a_{\pmb{q}}=f(\pmb{q})M(\pmb{q}),
\end{equation}
where $M(\pmb{q})$ is the scattering matrix element (for the acoustic phonons) and $f(\pmb{q})=\int{\rm d}\pmb{r}n_e(\pmb{r})e^{-i\pmb{q}\cdot\pmb{r}}$ is the form factor depending on the (equilibrium) charge density distribution $n_e(\pmb{r})=\sum_i^{\rm occ.} |\Psi_i(\pmb{r},t\rightarrow-\infty)|^2$ in the ring.\\
Assuming linear acoustic phonons with a dispersion $\omega_{\pmb{q}}=s|\pmb{q}|$ where $s$ is the sound velocity, we consider only the piezoelectric phonons and neglect contribution from the deformation potential which is justifiable at temperatures below 10\,K for bulk GaAs or InAs material \cite{brandes2002adiabatic, stavrou2005charge}. Therefore, $M(\pmb{q})^2=g_{\rm pz}\pi^2s^2/(V|\pmb{q}|)$ where $V$ is the volume of the unit cell and $g_{\rm pz}$ the dimensionless piezoelectric constant ($g_{\rm pz}=0.45$ for InAs \cite{brandes1999spontaneous}). We evaluated the spectral density function $J(\omega)$ numerically by transforming the sum in Eq.\,\eqref{eq:Jw} into an integral. By fitting we found that the high-frequency tail falls off like $J(\omega\rightarrow\infty)\varpropto1/\omega$ (Ohmic behavior). The bath correlation function is obtained from Eq.\,\eqref{eq:correlation} and a subsequent (numerically performed) Fourier transform leads to the bath-dependent parts of the Redfield components in Eq.\,\eqref{eq:RedfieldComponents} where we also need the scattering matrix elements $\langle i|\hat{Q}_{\pmb{\nu}}|j\rangle$ of the electron-phonon interaction between the various spin-resolved electronic states of the quantum ring.

\section{Dynamics by optical vortex pulse}
\label{AppB}

The typical time scale of the acoustic phonon-induced relaxation processes in III-V  semiconductor nanostructures is in the range of $>10$\,ps \cite{murdin1996time}, much longer than the considered few-cycle vortex pulse length $T_p$ . Therefore, during the action of the pulse, the relaxation is marginal.\\
An equilibrium electronic state characterized by the wave function $\Psi_i(\pmb{r})$ and the energy $E_i$ (note that we used the shortened notation $i=\left\{n,j,(\pm)\right\}$) can be decomposed in
\begin{equation}
\Psi_{n,j}^{(\pm)}(\pmb{r})  = \sum_{n_{\upar}=0}^{n_{\rm max}}a_{n_{\upar},\ell}^{n,(\pm)}|n_{\upar},\ell\rangle|\upar\rangle + \sum_{n_{\doar}=0}^{n_{\rm max}}b_{n_{\doar},\ell}^{n,(\pm)}|n_{\doar},\ell+1\rangle|\doar\rangle,
\label{eq:SObasis}
\end{equation}
where the (orbital) sub-states are given by $\langle\pmb{r}|n,\ell\rangle=R_{n,\ell}(r)e^{i\ell\varphi}$ with the radial wave functions $R_{n,\ell}(r)$ and angular quantum number $\ell$ ($\varphi$ is the polar angle). \\
In the presence of the optical vortex, the various (sub)states $|n,\ell\rangle|s\rangle$ are coupled due to different mechanisms. First, the direct (light-matter) interaction as part of $\hat{H}_{\rm int}$ with the optical vortex pulse is given by $\hat{H}_D=\frac{e}{2m^*}\left[\hat{\pmb{p}}\cdot\pmb{A}(\pmb{r},t) + \pmb{A}(\pmb{r},t)\cdot\hat{\pmb{p}}\right]$ and acts on the orbital part of the electron state. Thus, this Hamiltonian leaves the spin orientation $|s\rangle$ corresponding to the sub-states $|n,\ell\rangle$ (with coefficients $a_{n_{\upar},\ell}^{n,(\pm)}$ and $b_{n_{\doar},\ell}^{n,(\pm)}$) unaffected. The orbital coupling of the (orbital) sub-states is characterized by the (transition) matrix elements $\mathcal{M}_{n_j,n_i}^{\ell_j,\ell_i}=\langle n_j\ell_j|\hat{H}_D|n_i\ell_i\rangle$ which read explicitly
\begin{equation}
\begin{split}
\mathcal{M}_{n_j,n_i}^{\ell_j,\ell_i}=&\frac{ieA_0\hbar}{m_e}\sum_{\lambda=\pm1}\left[
\frac{2}{w_0^3}u_{n_j,n_i}^{\ell_j,\ell_i+\lambda}\right. \\
&-\frac{1}{w_0}\left(m_{\rm OAM}(1+\lambda)+2\lambda\ell_i-1\right)v_{n_j,n_i}^{\ell_j,\ell_i+\lambda} \\
&\left.-\frac{2}{w_0}w_{n_j,n_i}^{\ell_j,\ell_i+\lambda}\right]\delta_{\ell_j-\ell_i,m_{\rm OAM}+\lambda}.
\end{split}
\label{eq:HD1}
\end{equation}
The summation over $\lambda$ stems from the linear polarization of the optical vortex in the $x$-direction. The $\delta_{n,n'}$ is the Kronecker symbol while the coefficients $u,v$ and $w$ are radial integrals (in polar coordinates):
\begin{equation}
\begin{split}
u_{n',n}^{\ell',\ell}&=\int_0^\infty{\rm d}r\,rR_{n',\ell'}(r)R_{n,\ell}(r)f_{m_{\rm OAM}}(r), \\ v_{n',n}^{\ell',\ell}&=\int_0^\infty{\rm d}r\, R_{n',\ell'}(r)R_{n,\ell}(r)f_{m_{\rm OAM}}(r)/r~~\text{and } \\ w_{n',n}^{\ell',\ell}&=\int_0^\infty{\rm d}r\, R_{n',\ell'}(r)\partial_r[R_{n,\ell}(r)] f_{m_{\rm OAM}}(r).
\end{split}
\end{equation}
Note that $f_{m_{\rm OAM}}(r)$ is the radial distribution function of the optical vortex beam which dependents on the topological charge $m_{\rm OAM}$. Noticeably, the light-matter interaction leads to transitions from $|\Psi_n\rangle$ to quantum states $|\Psi_{n'}\rangle$ with higher orbital angular momenta as dictated by the winding number $m_{\rm OAM}$. Interestingly, although $\hat{H}_{\rm D}(t)$ does not act directly on the spin-orientation of the sub-states, a spin-flip transition is nevertheless possible due to the orbital coupling between sub-states of the two different (spin) channels $(+)$ and $(-)$.\\
The SOI interaction $\hat{H}_{\rm SOI}(t)=\frac{\alpha}{\hbar}\left[\hat{\sigma}\times\pmb{A}(\pmb{r},t)\right]_z$ describes the coupling of the spin state with the vector potential of the vortex beam. Since we consider a linearly polarized light in $x$-direction, the cross product is $\left[\hat{\sigma}\times\hat{\epsilon}\right]_z=-\sigma_y$. The photo-induced SOI transition matrix elements $\mathcal{S}_{n_j,n_i}^{\ell_j,\ell_i,s_j,s_i}=\langle s_j|\langle n_j\ell_j|\hat{H}_{\rm SOI}|n_i\ell_i\rangle|s_i\rangle=-\frac{\alpha}{\hbar}\langle n_j\ell_j|A_x(\pmb{r},t)|n_i\ell_i\rangle\langle s_j|\sigma_y|s_i\rangle$ is given by
\begin{equation}
\begin{split}
\mathcal{S}_{n_j,n_i}^{\ell_j,\ell_i,s_j,s_i}=s_i\widetilde{\alpha}\,h_{n_j,n_i}^{\ell_j,\ell_i}\delta_{\ell_j-\ell_i,m_{\rm OAM}}\delta_{s_j,-s_i},
\label{eq:HSOI1}
\end{split}
\end{equation}
where $\widetilde{\alpha}=iA_0e\alpha_R/\hbar$ and the radial integral is $h_{n,n'}^{\ell,\ell'}=\int_0^{\infty}{\rm d}r\,R_{n'\ell'}(r)R_{n\ell}(r)f_{m_{\rm OAM}}(r)$. This laser-induced interaction leads to spin-flip transitions within one spin channel as well as between the two different channels $(+)$ and $(-)$ due to $\langle s_j|\sigma_y|s_i\rangle$ (in the sub-states). Further, the (internal) orbital angular momentum state is changed due transfer of orbital angular momentum.\\
The electric scalar potential satisfies the Lorenz gauge conditions $\Phi(\pmb{r},t)=-c^2\int_{-\infty}^t{\rm d}t'\,\nabla\cdot\pmb{A}(\pmb{r},t')$ and gives rise to matrix elements of the form $\mathcal{K}_{n_j,n_i}^{\ell_j,\ell_i}=\langle n_j\ell_j|H_{\rm el}|n_i\ell_i\rangle$ (the spin orientation is again conserved):
\begin{equation}
\begin{split}
\mathcal{K}_{n_j,n_i}^{\ell_j,\ell_i}=&\frac{2eA_0}{\omega_xc^2}\sum_\lambda\left[\frac{1}{w_0^3}u_{n_j,n_i}^{\ell_j,\ell_i+\lambda} - \frac{\lambda m_{\rm OAM}}{w_0}v_{n,n'}^{\ell,\ell'+\lambda}\right] \\
&\times\delta_{\ell_j-\ell_i,m_{\rm OAM}+\lambda}.\\
~
\end{split}
\end{equation}
This light-induced transition initiated by the electric scalar potential $\Phi(\pmb{r},t)$ is comparable to the direct interaction $\hat{H}_D$ in Eq.\,\eqref{eq:HD1}. Thus, indirect spin-flip transitions are possible due to orbital couplings between both spin channels. Furthermore, the internal orbital angular momentum state changes during the interaction.

%\bibliography{v1}
